\documentclass[sigconf]{acmart}
\renewcommand\footnotetextcopyrightpermission[1]{} 
\AtBeginDocument{%
  }

\acmConference{Preprint submitted to arXiv}{}

\begin{document}

\title{Accelerating Learned Video Compression via Low-Resolution Representation Learning}


\author{Zidian Qiu}

\author{Zongyao He}

\author{Zhi Jin}





\renewcommand{\shortauthors}{Qiu et al.}


\begin{abstract}
In recent years, the field of learned video compression has witnessed rapid advancement, exemplified by the latest neural video codecs DCVC-DC that has outperformed the upcoming next-generation codec ECM in terms of compression ratio. Despite this, learned video compression frameworks often exhibit low encoding and decoding speeds primarily due to their increased computational complexity and unnecessary high-resolution spatial operations, which hugely hinder their applications in reality. In this work, we introduce an efficiency-optimized framework for learned video compression that focuses on low-resolution representation learning, aiming to significantly enhance the encoding and decoding speeds. Firstly, we diminish the computational load by reducing the resolution of inter-frame propagated features obtained from reused features of decoded frames, including I-frames. We implement a joint training strategy for both the I-frame and P-frame models, further improving the compression ratio. Secondly, our approach efficiently leverages multi-frame priors for parameter prediction, minimizing computation at the decoding end. Thirdly, we revisit the application of the Online Encoder Update (OEU) strategy for high-resolution sequences, achieving notable improvements in compression ratio without compromising decoding efficiency. Our efficiency-optimized framework has significantly improved the balance between compression ratio and speed for learned video compression. In comparison to traditional codecs, our method achieves performance levels on par with the low-decay P configuration of the H.266 reference software VTM. Furthermore, when contrasted with DCVC-HEM, our approach delivers a comparable compression ratio while boosting encoding and decoding speeds by a factor of 3 and 7, respectively. On RTX 2080Ti, our method can decode each 1080p frame under 100ms. 
\end{abstract}



\maketitle

\section{Introduction}
Video coding is a crucial technique aimed at reducing the size of video files by compressing redundant information intra and inter video frames. Traditional video coding standards such as H.264/AVC \cite{h264}, H.265/HEVC \cite{h265}, AV1 \cite{av1}, AVS3 \cite{avs3}, and H.266/VVC \cite{h266} employ a series of complex algorithms and techniques to compress redundant video data and reduce file sizes while providing high-quality decoded videos. These standards strive to offer higher compression efficiency and better video quality to meet the growing demands for video content. With the rise of deep learning, research on end-to-end learned video compression frameworks has yielded promising results in recent years. Some of these works have achieved compression ratios surpassing the latest video coding standard H.266. Nevertheless, the high complexity of learned video compression methods becomes the primary factor that reduces encoding and decoding speeds, and limits their applications in reality. Therefore, optimizing learned video compression is a significant challenge, necessitating a reevaluation of framework designs.

\begin{figure}[t]
  \includegraphics[width=\linewidth]{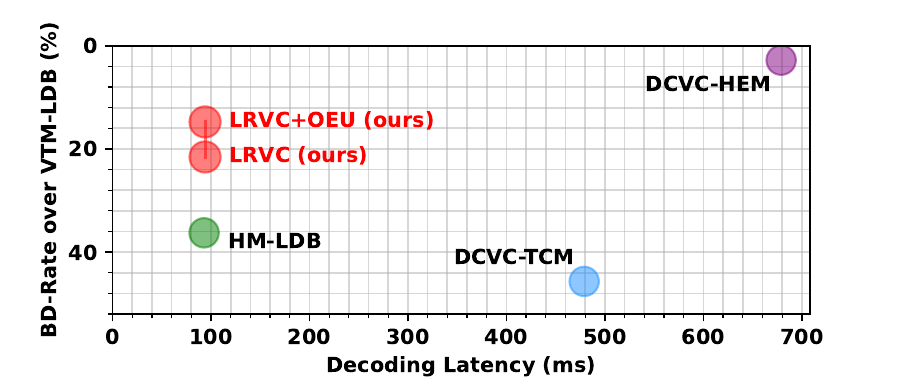}
  \caption{Rate-speed comparison on UVG dataset. Methods closer to the top left provide better performance and efficiency. OEU represents the online encoder update strategy. Tested on RTX 2080Ti using 1080p as the input.}
  \label{fig:teaser}
\end{figure}

In end-to-end neural video codecs, all critical components of video compression, e.g., motion estimation, motion compensation, motion compression, and residual compression, are implemented through end-to-end neural networks, enabling joint optimization. Present end-to-end works can generally be categorized into two main categories: residual coding and conditional coding. In residual coding, predicted frames are first generated from previously decoded frames, following which the residual between the current frame and the predicted frame is computed. This residual is encoded into a bitstream and decoded to obtain the reconstructed residual, which is then added to the predicted frame to derive the final decoded frame. DCVC \cite{dcvc} proposed a conditional coding framework that uses features propagated between frames as context inputs to the contextual encoder and decoder. Notably, DCVC-DC \cite{dcvc_dc} is the first end-to-end neural video codec to achieve the compression ratio surpassing the next-generation codec prototype ECM \cite{ecm}, highlighting the leading advantage of learned video compression. However, its drawbacks are pronounced, as on an RTX 2080Ti, it requires more than one second to encode and decode a 1080p frame, posing significant challenges for the practical deployment of neural video codecs.

\begin{figure*}[ht]
  \centering
  \includegraphics[width=0.9\textwidth]{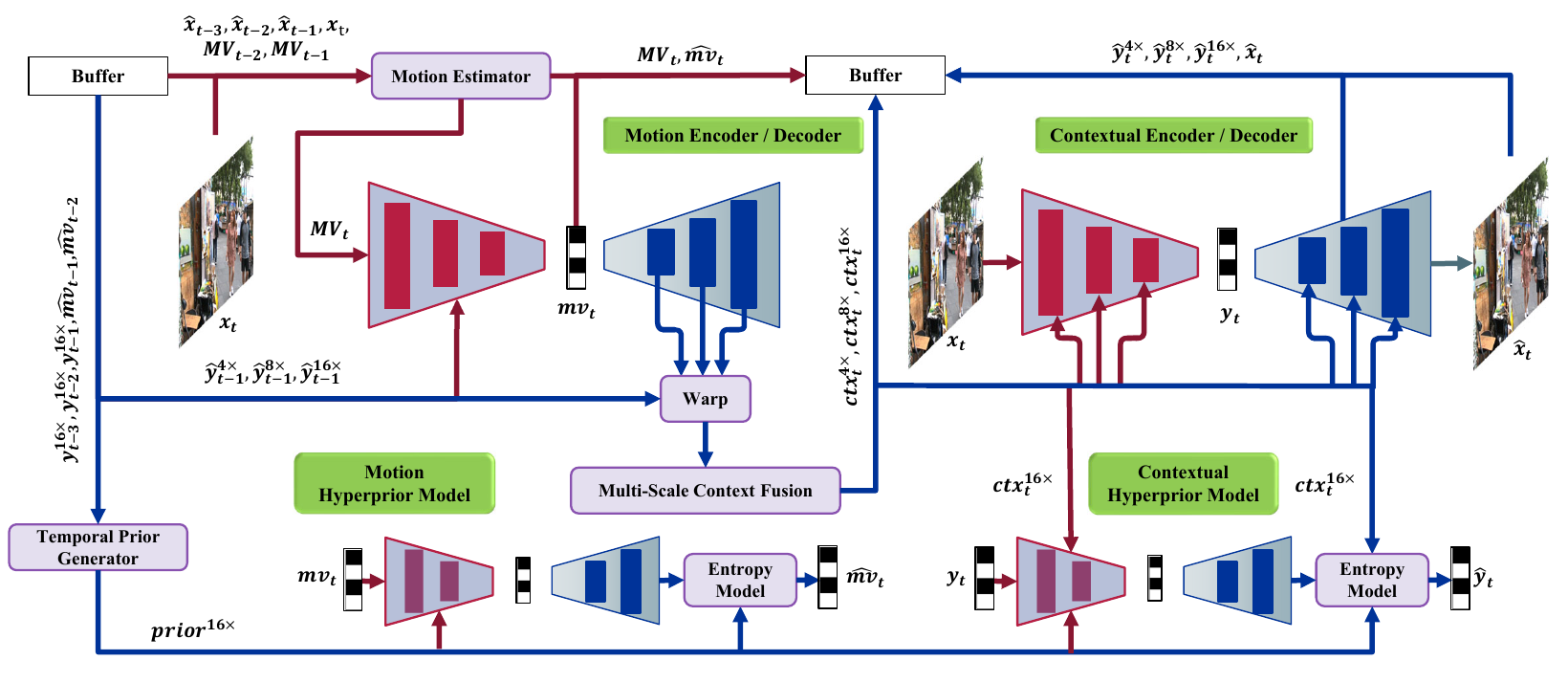}
  \caption{Overview of our framework, where red lines represent data flows not included in the decoder side, while blue lines indicate data flows on the decoder side. $x_t$ and $\hat{x}_t$ represent the input frame and reconstructed frame, respectively. $MV_t$, $mv_t$ and $\hat{mv}_t$ denote the optical flow, motion vector and reconstructed motion vector, respectively. $y_t$ and $\hat{y}_t$ represent the latent representation and reconstructed intermediate feature in the contextual decoder, and $ctx_t$ is the learned temporal context, where the superscript $n\times$ denotes that the downsampling factor is $n$ relative to the input resolution.}
  \label{fig2}
\end{figure*}

Optimizing the conditional coding framework poses significant challenges in terms of inference latency due to the multitude of modules involved. As a solution, in this work, we introduce an efficiency-optimized framework for learned video compression that focuses on low-resolution representation learning, aiming to significantly enhance encoding and decoding speeds, as shown in Figure \ref{fig2}. Firstly, we identify that the majority of high-resolution spatial operations contribute significantly to inference latency. To mitigate this, we strive to minimize high-resolution operations by relocating them to low-resolution space wherever possible. Additionally, we implement feature reuse for decoded frames to save on extra computational costs. Interestingly, we find that the I-frame model can also benefit from feature reuse, and can be jointly optimized with the P-frame model in our framework. While some previous works \cite{prior1,prior2,prior3,prior4,rippel2021elf,canf_vc} explore the utilization of multi-frame priors, most of them require high-resolution space execution at the decoding end. In contrast, we employ high-quality priors at the encoding end and generate temporal priors from the low-resolution feature domain at the decoding end, resulting in negligible additional computational overhead. Leveraging multi-frame priors for entropy model parameter prediction improves performance compared to single-frame priors. Furthermore, we revisit the application of the Online Encoder Update (OEU) strategy \cite{oeu} on high-resolution sequences, further enhancing performance without affecting decoding speed. Our method achieves significantly faster decoding time compared to other neural video codecs at similar compression ratios, as shown in Figure \ref{fig:teaser}. Compared to traditional codecs, our method performs on par with the low-decay P configuration of H.266 reference software VTM \cite{h266}. Although our method remains a gap in PSNR compared to the low-decay B configuration of VTM, the MS-SSIM \cite{msssim} results are better. In summary, our contributions are outlined as follows:
\begin{itemize}
\item We identify that a major source of latency arises from high-resolution spatial operations. To mitigate this, we relocating them to low-resolution spaces. We introduce feature reuse in decoding frames (including I-frames) to reduce additional computational costs, and jointly optimize the I-frame model with the P-frame model.
\item By generating temporal priors in the low-resolution feature domain at the decoding end, we utilize multi-frame priors for entropy model parameter prediction to improve performance with negligible additional computational overhead. By revisiting the application of the OEU strategy, we further improve performance without compromising decoding speed. 
\item Compared to traditional codecs, our method achieves comparable performance to the low-decay P configuration of VTM \cite{h266}. In comparison to DCVC-HEM \cite{dcvc_hem}, our approach achieves comparable compression ratio while increasing encoding and decoding speeds by $3\times$ and $7\times$.
\end{itemize}

\section{Related Work}

\subsection{Traditional Video Compression}
The manual design of video codecs has a long history. Traditional video codecs mainly refer to techniques based on coding standards, such as H.264/AVC \cite{h264}, H.265/HEVC \cite{h265}, AV1 \cite{av1}, AVS3 \cite{avs3}, H.266/VVC \cite{h266}, etc. These standards aim to provide higher compression efficiency and better video quality to meet the ever-growing demand for videos. However, they are meticulously designed and tuned, employing a series of complex algorithms and technologies, and
cannot be jointly optimized in an end-to-end way.

\subsection{Learned Video Compression}
DVC \cite{lu2019dvc}, proposed by Lu\textit{ et al.}, was the first end-to-end learned video compression framework utilizing deep learning. In FVC \cite{hu2021fvc}, Hu\textit{ et al.} improves compression ratio by operating all major operations in the feature domain. Hu\textit{ et al.} further improves compression ratio by introducing coarse-to-fine motion estimation and incorporating hyperprior-guided adaptive motion compression and hyperprior-guided adaptive residual compression in C2F \cite{c2f}. MMVC \cite{liu2023mmvc} adapts different strategies for frames of different types based on encoded residuals to accommodate various motion patterns present on different blocks within intra-frames. Li\textit{ et al.} proposed DCVC \cite{dcvc} to devise an efficient conditional coding framework, utilizing temporal context features as conditional inputs to assist the codec in encoding and decoding the current frame. DCVC-TCM \cite{dcvc_tcm} adopts multi-scale temporal contexts to boost performance. DCVC-HEM \cite{dcvc_hem} designs a sophisticated entropy model, being the first to deploy an end-to-end neural video codec surpassing the highest compression ratio configuration of the H.266 reference software VTM. DCVC-DC \cite{dcvc_dc} uses the group-based offset diversity to enhance optical flow-based coding framework, and provides diverse spatial contexts for better entropy coding through quadtree-based partitioning, being the first end-to-end neural video codec surpassing the highest compression ratio configuration of the next-generation codec prototype ECM \cite{ecm}. Although neural video codecs have caught up with or even surpassed traditional codecs, they still lag behind traditional video codecs in the trade-off between compression ratio and computational efficiency.

\subsection{Efficient Learned Video Compression}
To date, only a few studies have focused on designing computationally efficient video codecs. ELF-VC \cite{rippel2021elf} adopts a specially optimized backbone network that achieves robust performance in video compression while maintaining computational efficiency. AlphaVC \cite{shi2022alphavc} uses an efficient probability-based entropy skipping strategy that significantly reduces the computational load of arithmetic coding at both the encoding and decoding ends. MobileCodec \cite{le2022mobilecodec} stands out as the first neural video codec to run on commercial smartphones, using quantization-aware training to quantize network parameters and activations to fixed point and developing a parallel entropy coding algorithm. Tian\textit{ et al.} \cite{mm2023} has developed a lightweight model leveraging a series of efficiency techniques such as model pruning, motion subsampling, and entropy skipping, addressing potential inaccuracies in bitstream decoding arising from cross-platform computation errors caused by floating-point operations. However, the trade-off for these methods in optimizing inference speed is a significant performance degradation, resulting in low compression ratios. For instance, Tian\textit{ et al.} proposed a method that achieved real-time decoding speed on 720p sequences but exhibited performance comparable only to the medium preset of x265. In contrast, we introduce an efficiency-optimized framework that achieves over 50\% bitrate savings compared to Tian\textit{ et al.} while maintaining comparable decoding speed.

\section{Method}

\subsection{Overall Framework}

\begin{figure}[h]
  \includegraphics[width=\linewidth]{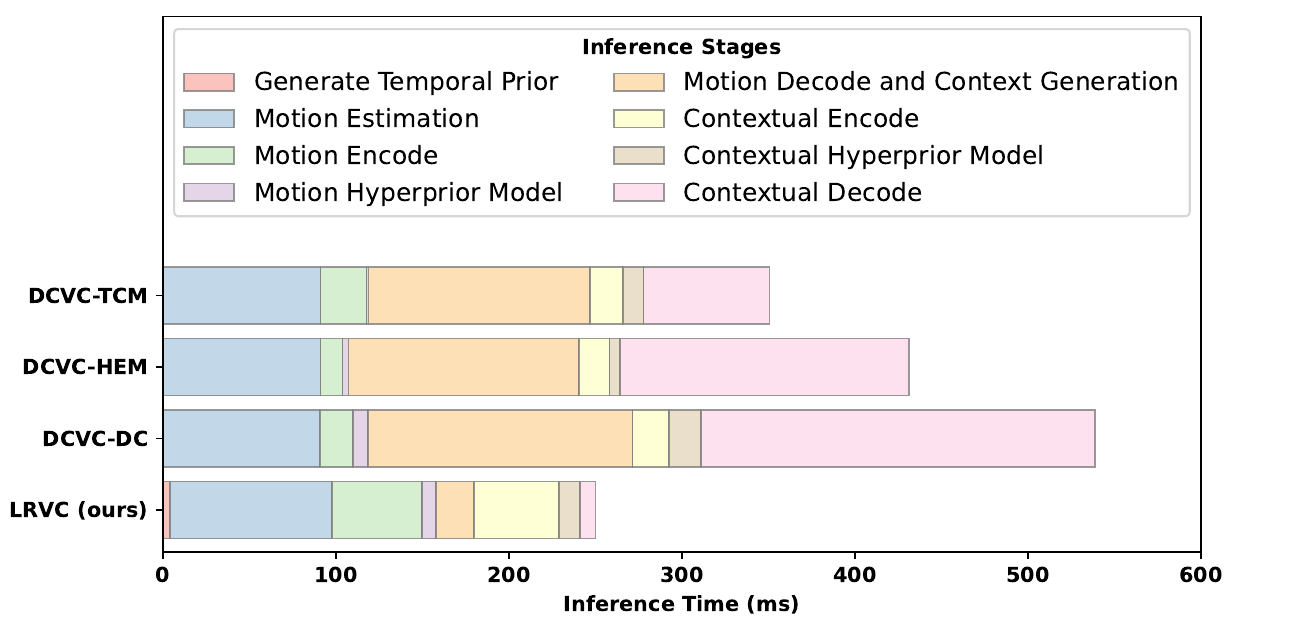}
  \caption{Inference latency composition of different approaches. Tested on V100 using 1080p as the input.}
  \label{fig3}
\end{figure}

Given our focus on optimizing network architecture from the perspective of actual inference latency, we begin by designing the network using simple residual blocks and residual bottleneck blocks \cite{resnet} and minimize the usage of depth-wise convolutions \cite{howard2017mobilenets} that is unfriendly to GPU inference. For the majority of conventional online video applications, encoding can typically be conducted offline, thus placing greater emphasis on decoding speed. Accordingly, we adopt an imbalanced design approach, prioritizing complexity optimization at the decoding end while using higher-complexity networks for the encoder. By optimizing the entire framework, we significantly improve decoding speed, as shown in Figure \ref{fig3}. In our framework, for the I-frame model, the encoder consists of stacked residual blocks, while the decoder consists of stacked residual bottleneck blocks to reduce computational overhead. The P-frame framework is based on the conditional coding framework \cite{dcvc_tcm,dcvc_hem,dcvc_dc}, as shown in Figure \ref{fig2}. The main modules are introduced below.

\begin{figure}[t]
  \includegraphics[width=\linewidth]{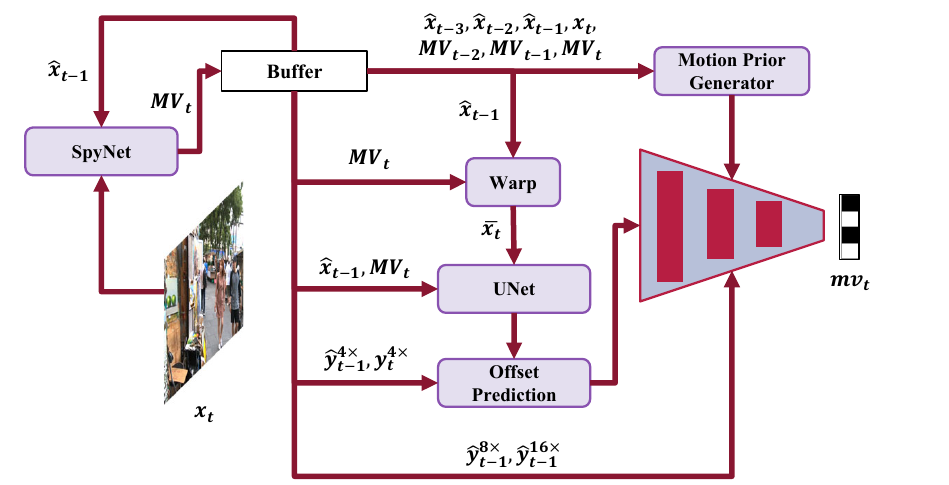}
  \caption{Proposed motion encoder utilizing multi-frame priors.}
  \label{fig4}
\end{figure}

\textbf{Motion Estimation and Motion Encoding.} For the input frame, we first estimate optical flow $MV_t$ between the current frame $x_t$ and the reference frame $\hat{x}_{t-1}$ using the SpyNet \cite{spynet} optical flow estimation network. We then refine the motion feature using a UNet \cite{unet} with the warped frame $\overline{x}_{t}$ and optical flow as input. Additionally, we perform feature-level motion estimation to generate the offsets, which are then fused with the previously obtained motion feature. Furthermore, we extract temporal priors from additional reference frames and optical flows ($\hat{x}_{t-3}$, $\hat{x}_{t-2}$, $\hat{x}_{t-1}$, $MV_{t-2}$, $MV_{t-1}$) to serve as input to the motion encoder and generate the motion vector $mv_t$, as shown in Figure \ref{fig4}.

\textbf{Motion Decoding and Temporal Context Alignment.} After motion encoding, we decode the compressed motion vector progressively. For each resolution, the decoded features serve as offsets to align the features of the reference frame through deformable alignment. The aligned features are then input to the multi-scale context fusion module to augment the contextual information.

\textbf{Contextual Encoder-Decoder.} During encoding, we conditionally encode the current frame into a latent representation using multi-scale contexts ($ctx_t^{4\times}, ctx_t^{8\times}, ctx_t^{16\times}$). During decoding, we similarly use the multi-scale contexts to reconstruct the latent representation $\hat{y}_t$ into a reconstructed frame $\hat{x}_t$. To alleviate the burden on the decoding end, we employ a lightweight decoder without any post-processing operations. To compensate for the performance loss incurred by the lightweight decoder, we leverage an imbalanced design approach by scaling up the number of layers and channels at the encoding end.

\textbf{Hyperprior Model and Entropy Model.} We use Laplace distribution to model the motion vector and latent representation in the motion hyperprior model and contextual hyperprior model. Both our I-frame model and P-frame model utilize arithmetic coding skip entropy models \cite{shi2022alphavc} which are based on quadtree \cite{dcvc_dc}. To efficiently utilize multi-frame priors at the decoding end, we employ stacked residual bottleneck blocks to generate temporal priors from the latent representation in the decoding buffer, serving as inputs to the motion hyperprior model and contextual hyperprior model for more accurate parameter prediction.

\subsection{Low-Resolution Representation Learning}

Taking DCVC series \cite{dcvc_tcm,dcvc_hem,dcvc_dc} as an example, Figure \ref{fig3} illustrates the composition of their inference latency, where the decoding latency constitutes the majority. Through the analysis of their inference latency, we identify that a significant portion of the decoding latency originates from high-resolution spatial operations, such as the context fusion module for high-resolution context and the post-processing stage of the contextual decoder. What if we neglect the optimization of these high-resolution spatial operations? For example, in Section \ref{sec:abstudy}, we attempted to accelerate DCVC-DC by reducing model depth and channel numbers. The experimental results reveal an unsatisfactory acceleration effect, and the model loses its ability to converge. Therefore, we optimize these high-resolution spatial operations through low-resolution representation learning.

In contrast to the high-resolution multi-scale contexts used in DCVC-TCM \cite{dcvc_tcm}, we employ low-resolution multi-scale contexts to reduce subsequent high-resolution spatial operations. We first reduce the resolution of the contexts, making them $4\times$, $8\times$, and $16\times$ downsampled compared to the original resolution, respectively. They are obtained from the intermediate features from the I-frame decoder or P-frame decoder of the previous time step and dimensionally reduced to 64, 64, and 128 through feature adaptors. In previous works \cite{dcvc_tcm,dcvc_hem,dcvc_dc}, context features are obtained by convolutional downsampling from the previous decoded frame, with downsampling factors of $1$, $2$, and $4$ compared to the original resolution, which involves a considerable amount of high-resolution spatial operations and incurs high computational costs. In contrast, our method for obtaining context features does not require additional computational costs and only requires reusing decoded features.

\begin{figure}[t]
  \includegraphics[width=0.7\linewidth]{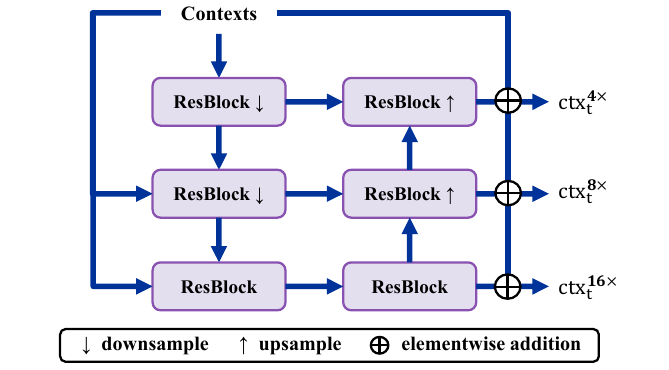}
  \caption{Proposed multi-scale context fusion module.}
  \label{fig5}
\end{figure}

After reducing the resolution of temporal contexts, we need to reconsider the design of the motion decoder. Previous works \cite{dcvc_tcm,dcvc_hem,dcvc_dc} use optical flow to align multi-scale contexts. However, we found that the performance of optical flow alignment deteriorates significantly when the temporal context resolution is reduced. Therefore, we employ deformable convolution for alignment \cite{dcn}. Our motion decoder performs motion decoding and deformable alignment simultaneously. Since motion decoding and motion compensation are performed in the feature domain using deformable convolution, without involving high-resolution operations, we avoid the need to decode to the original resolution, as required by optical flow decoding, thereby circumventing high-resolution spatial operations. After generating multi-scale contexts, we input them into the context fusion module shown in Figure \ref{fig5} for multi-scale fusion. Due to the reduction in temporal context resolution, this operation is also performed in low-resolution space.

\subsection{Exploring Richer Prior Information}\ 

Maximizing the utilization of prior information is a crucial topic explored in previous works. In video coding, prior information often refers to the effective information contained in the decoded frames, such as optical flows from previous frames and intermediate features. Previous works have employed techniques like motion extrapolation, residual prediction, motion and residual refinement to improve compression ratio \cite{prior1,prior2,prior3,prior4,rippel2021elf,canf_vc}. These operations need to be performed at the decoding end and operate at high resolution, significantly increasing decoding complexity and latency. However, compared to their increased computational costs, these methods have a limited impact on improving the compression ratio.

To address this challenge, we enhance temporal priors by increasing the number of reference frames, leveraging information from more distant frames to model the current frame. At the encoding end, we utilize high-quality priors, i.e., reference frames and optical flows at original resolution, to perform more accurate motion estimation, as illustrated in Figure \ref{fig4}. Meanwhile, at the decoding end, we generate temporal priors solely from the low-resolution feature domain. These priors are then used as inputs to the motion hyperprior model and contextual hyperprior model for more accurate parameter prediction. Thus, by introducing negligible additional computational overhead at the decoding end, we exploit multi-frame priors, which leads to performance improvements compared to utilizing single-frame priors.

By directly utilizing intermediate features generated during the decoding process of the reference frame, we avoid additional computations and minimize information loss. Particularly, since the I-frame generally serves as a high-quality reference frame, fully utilizing the high-quality information contained in the I-frame is pivotal. We observe that prior works \cite{lu2019dvc,canf_vc,dcvc_tcm,dcvc_hem,dcvc_dc} often overlook the significance of the I-frame model, resorting to traditional encoder or training independent I-frame model for encoding. Within a conditional coding framework, the temporal contexts for the first P-frame are obtained by extracting features from the decoded I-frame, hindering the full utilization of high-quality information in the I-frame. Similar to the P-frame model, we enhance the coupling between the I-frame model and P-frame model by reusing intermediate features from the decoding process of the I-frame, thereby strengthening both the I-frame and P-frame models and achieving further performance improvements through joint training.

\begin{table*}[t]
\centering
\caption{Comparison of model parameters, MACs, encoding time and decoding time with 1080p resolution input.}
\renewcommand{\arraystretch}{1.25}
\small
\begin{tabular}{cccccc}
\toprule[1.0pt]
Model & \#Params(I model/P Enc/P Dec) & MACs(Enc/Dec) & Encode time & Decode time & Device \\ \hline
HM & - & - & 42s & 93ms & AMD EPYC 7352 \\
VTM & - & - & 189s & 132ms & AMD EPYC 7352 \\ \hline
DCVC-TCM \cite{dcvc_tcm} & 23.7M/\textbf{10.7M}/\textbf{7.5M} & 2.960T/1.908T & 632ms & 341ms & V100 \\
DCVC-HEM \cite{dcvc_hem} & 31.2M/17.5M/15.3M & 3.465T/2.586T & 595ms & 468ms & V100 \\
DCVC-DC \cite{dcvc_dc} & 31.0M/19.8M/15.7M & 2.790T/1.901T & 603ms & 477ms & V100 \\
LRVC (ours) & \textbf{12.5M}/35.1M/10.6M & \textbf{1.824T}/\textbf{0.277T} & \textbf{257ms} & \textbf{86ms} & V100 \\ \hline
DCVC-TCM \cite{dcvc_tcm} & 23.7M/\textbf{10.7M}/\textbf{7.5M} & 2.960T/1.908T & 892ms & 479ms & RTX 2080Ti \\
DCVC-HEM \cite{dcvc_hem} & 31.2M/17.5M/15.3M & 3.465T/2.586T & 974ms & 679ms & RTX 2080Ti \\
DCVC-DC \cite{dcvc_dc} & 31.0M/19.8M/15.7M & 2.790T/1.901T & 948ms & 697ms & RTX 2080Ti \\
LRVC (ours) & \textbf{12.5M}/35.1M/10.6M & \textbf{1.824T}/\textbf{0.277T} & \textbf{383ms} & \textbf{94ms} & RTX 2080Ti \\
\bottomrule[1.0pt]
\end{tabular}%
\label{tab_model_comparison}%
\end{table*}%

\begin{table*}[t]
\centering
\caption{BD-Rate (\%) comparison for PSNR on UVG, MCL-JCV and HEVC B datasets. The anchor is VTM-LDB.}
\renewcommand{\arraystretch}{1.25}
\small
\resizebox{\linewidth}{!}{
\begin{tabular}{cccccccccc}
\toprule[1.0pt]
& VTM-LDB & HM-LDB & DCVC \cite{dcvc} & CANF-VC \cite{canf_vc} & DCVC-TCM \cite{dcvc_tcm} & DCVC-HEM \cite{dcvc_hem} & DCVC-DC \cite{dcvc_dc} & LRVC (ours) & LRVC+OEU (ours) \\ \hline
UVG & 0 & 36.26 & 155.01 & 64.54 & 45.68 & 2.79 & \textbf{-17.98} & 21.55 & 14.76 \\ \hline
MCL-JCV & 0 & 41.78 & 114.07 & 65.34 & 50.85 & 8.74 & \textbf{-10.53} & 30.56 & 18.99 \\ \hline
HEVC B & 0 & 39.06 & 117.08 & 60.49 & 40.36 & 5.08 & \textbf{-12.04} & 32.07 & 22.93 \\
\bottomrule[1.0pt]
\end{tabular}}
\label{tab1}%
\end{table*}%

\begin{table*}[t]
\centering
\caption{BD-Rate (\%) comparison for MS-SSIM on UVG, MCL-JCV and HEVC B datasets. The anchor is VTM-LDB.}
\renewcommand{\arraystretch}{1.25}
\small
\resizebox{\linewidth}{!}{
\begin{tabular}{cccccccccc}
\toprule[1.0pt]
& VTM-LDB & HM-LDB & DCVC \cite{dcvc} & CANF-VC \cite{canf_vc} & DCVC-TCM \cite{dcvc_tcm} & DCVC-HEM \cite{dcvc_hem} & DCVC-DC \cite{dcvc_dc} & LRVC (ours) & LRVC+OEU (ours) \\ \hline
UVG      & 0     & 27.26 & 58.79 & 39.86 & 0.03 & -25.11 & \textbf{-32.50} & -2.01 & -9.80 \\ \hline
MCL-JCV  & 0     & 35.59 & 26.00    & 22.43 & -10.58 & -36.12 & \textbf{-44.61} & -8.58 & -15.61 \\ \hline
HEVC B   & 0     & 38.27 & 54.47 & 42.80   & -11.18 & -37.65 & \textbf{-47.57} & -9.78 & -14.76 \\
\bottomrule[1.0pt]
\end{tabular}}
\label{tab2}%
\end{table*}%

\begin{table*}[h]
\centering
\caption{BD-Rate (\%) comparison for PSNR on UVG, MCL-JCV and HEVC B datasets. The anchor is VTM-LDP.}
\renewcommand{\arraystretch}{1.25}
\small
\resizebox{0.75\linewidth}{!}{
\begin{tabular}{ccccccccc}
\toprule[1.0pt]
& VTM-LDP & HM-LDP & DCVC \cite{dcvc} & ELF-VC \cite{rippel2021elf} & C2F \cite{c2f} & MMVC \cite{liu2023mmvc} & LRVC (ours) & LRVC+OEU (ours) \\ \hline
UVG & 0 & 33.24 & 95.87 & 65.03 & 16.79 & 2.53 & 4.91 & \textbf{-1.06} \\ \hline
MCL-JCV & 0 & 26.09 & 66.12 & 56.98 & 19.68 & \textbf{-8.56} & 5.29 & -1.87 \\ \hline
HEVC B & 0 & 45.17 & 67.36 & - & 14.19 & \textbf{-26.06} & 29.18 & 20.26 \\
\bottomrule[1.0pt]
\end{tabular}}
\label{tab3}%
\end{table*}%

\subsection{Online Encoder Update Strategy}
The online update strategy, distinguished for its adaptability, has garnered significant attention in various research domains \cite{bojanowski2017optimizing,park2019deepsdf,wang2022neural}. Applying this strategy to video coding frameworks can improve rate-distortion performance without increasing decoding time. Leveraging a learning-based approach, the network is guided through Rate-Distortion Optimization (RDO) during the training phase. However, due to domain gaps between training and testing samples, parameters obtained during training may not be optimal for the test set.

An Online Encoder Update (OEU) strategy tailored for video coding tasks is introduced in previous work \cite{oeu} to make the encoder content-adaptive. Nonetheless, a notable drawback of this method is its impracticality for high-resolution test sequences like 1080p sequences. This is primarily due to the requirement of feeding entire frames into the network for backpropagation after introducing error propagation-aware training mechanisms, leading to unacceptable memory consumption and limiting its practicality. Several schemes exist to address this issue, such as reducing the resolution of input sequences or conducting patch-wise training.

Through experiments, we found that simple overlapping patch-wise training can achieve satisfactory performance. Specifically, updates are only applied to the motion encoder, contextual encoder, and corresponding hyperprior encoders of the P-frame model. As we utilized weighted rate-distortion of $T$ frames as the loss during training, and the position within the Group Of Pictures (GOP) of each frame is fixed during testing, it has been found unnecessary to update all frames. Updating only the first $T$ frames of each GOP suffices to achieve impressive performance.



\section{Experiments}
\subsection{Implementation Details}

\textbf{\quad \  Dataset.} Our model is trained on the Vimeo-90k \cite{vimeo90k} dataset, which contains 89,800 video clips with a resolution of $448 \times 256$. During training, video frames are randomly cropped into $256 \times 256$ patches. We evaluate our model on UVG \cite{uvg}, MCL-JCV \cite{mcl_jcv}, and HEVC B \cite{hevc_datasets} datasets, all of which have a resolution of $1920 \times 1080$.

\textbf{Training Details.} We trained four models targeting different rate-distortion points with varying $\lambda$ values (85, 170, 380, 840), where $\lambda$ is the Lagrange multiplier that determines the trade-off between the number of bits and distortion. The loss function is a commonly used rate distortion loss. 
To stabilize the training process, we utilized gradient clipping and adopted the hierarchical quality structure training strategy proposed in DCVC-DC \cite{dcvc_dc}. Initially, we independently trained the I-frame model and the P-frame model. During the training process, we gradually increased the number of training frames from 2 to 7 before jointly training the I-frame model and P-frame models. 
For performance evaluation using MS-SSIM \cite{msssim}, we used $1 - \text{MS-SSIM}$ as the distortion loss for fine-tuning. 

\begin{figure*}[ht]
	\centering
	\includegraphics[width=\linewidth]{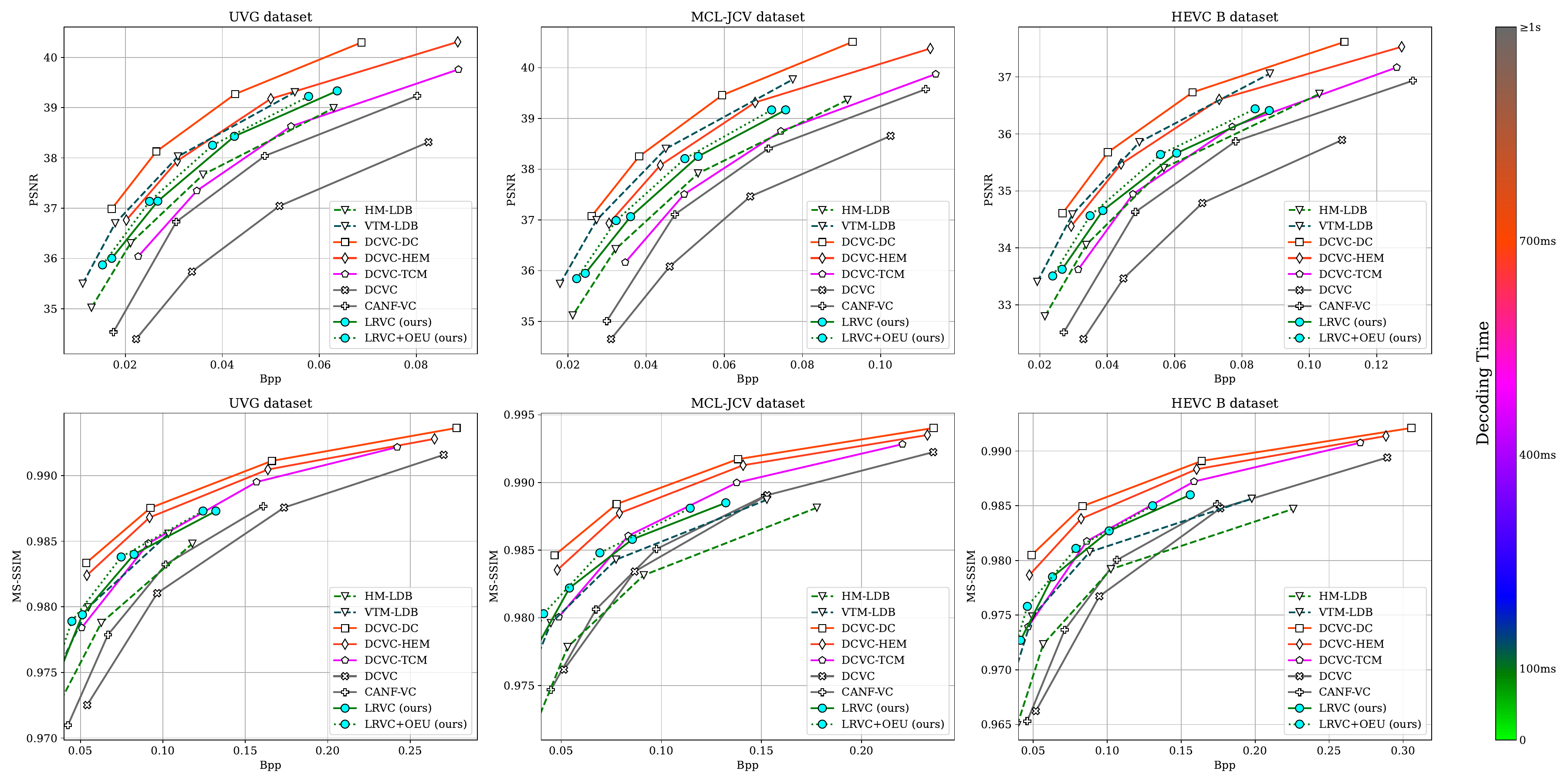} 
	\vspace{-0.8cm}
	\caption{Rate and distortion curves. The comparison is in RGB colorspace with BT.709 measured with PSNR and MS-SSIM. The line color represents the decoding time. 
	}
	\label{rd_curve12}
\end{figure*}

\begin{figure*}[t]
	\centering
	\includegraphics[width=\linewidth]{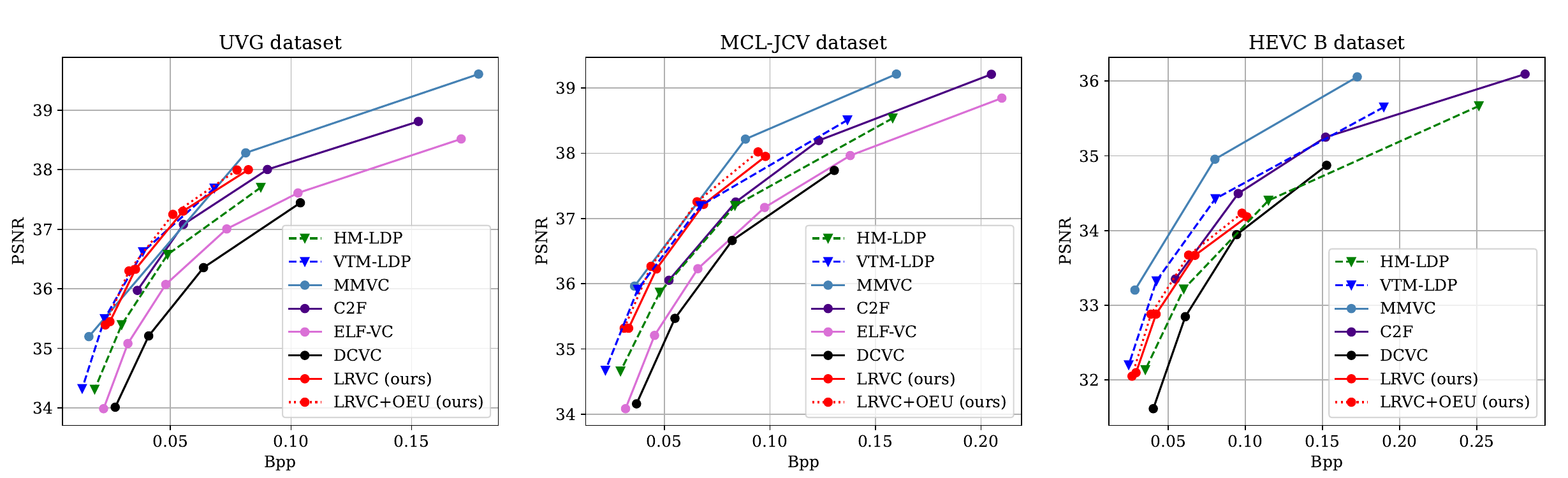} 
	\vspace{-0.8cm}
	\caption{Rate and distortion curves. The comparison is in RGB colorspace with BT.601 measured with PSNR.  
	}
	\label{rd_curve3}
\end{figure*}

\textbf{Test Conditions.} We evaluated the performance of all models using common metrics, such as PSNR and MS-SSIM and measured the compression ratio changes using BD-Rate \cite{bd_rate}, where positive values indicate bitrate increase and negative values indicate bitrate savings. To enable comprehensive comparisons, we assessed traditional codecs and neural video codecs in the RGB color space. For test sets originally in YUV420 format, we followed the settings of DCVC-DC, converting them to the RGB color space using BT.709. We tested the first 96 frames of each sequence with an intra-frame period of 32. To facilitate comparison, we referenced some results from DCVC-DC.

To compare our models under different test conditions with other models \cite{dcvc,rippel2021elf,c2f,liu2023mmvc}, we also tested our models following the test conditions of the PyTorchVideoCompression repository \footnote{See repository at \url{https://github.com/ZhihaoHu/PyTorchVideoCompression}}. This testing standard involved converting from YUV420 to the RGB color space using BT.601 (the relative bitrate comparisons between different codecs are similar in BT.601 and BT.709) and obtaining frames of $1920 \times 1024$ resolution by centrally cropping the original frames instead of padding to ensure the input frame shape is divisible by 64. The complete sequence was tested under these conditions.

\subsection{Comparison to Previous SOTA Methods}

Table \ref{tab_model_comparison} and Table \ref{tab1}, \ref{tab2}, \ref{tab3} present the efficiency analysis quantitative results of our method compared to other methods. Figure \ref{rd_curve12} and Figure \ref{rd_curve3} depict the corresponding rate-distortion curves. Our benchmark tests include HM \cite{h265} and VTM \cite{h266}, representing the best H.265 and H.266 codecs, as well as previous neural video codecs such as DCVC \cite{dcvc}, ELF-VC \cite{rippel2021elf}, CANF-VC \cite{canf_vc}, DCVC-TCM \cite{dcvc_tcm}, DCVC-HEM \cite{dcvc_hem}, DCVC-DC \cite{dcvc_dc}, C2F \cite{c2f}, and MMVC \cite{liu2023mmvc}.

\textbf{Efficiency Analysis.} In Table \ref{tab_model_comparison}, we analyze the efficiency of each model. Our proposed method comprises a total of 47.6 million parameters, including both the I-frame and P-frame models. Remarkably, the parameter count of our I-frame model is 12.5M, significantly smaller than the compared methods, while that of DCVC-TCM, DCVC-HEM and DCVC-DC is 23.7M, 31.2M and 31.0M. Although the parameter count of our P-frame model is relatively higher, our design philosophy of encoding and decoding imbalance results in a lower parameter count for our P-frame decoder. When the input sequence resolution is set to 1080p, the MACs (multiply-accumulate operations) for our method amount to 1.8T, with a decoding complexity of 277G. This is significantly lower compared to the other methods we compared against. Compared to DCVC-HEM, the encoding complexity and decoding complexity of DCVC-HEM are 1.9 and 9 times higher than ours, respectively. Notably, even though DCVC-DC utilizes parameter-efficient and computationally efficient depth-wise separable convolutions as basic blocks, its encoding complexity is nearly 1.6 times higher than ours, while the decoding complexity is approximately 7 times higher.

We also conducted tests to measure the average encoding and decoding times of different codecs, with the testing devices and results presented in Table \ref{tab_model_comparison}. Compared to traditional codecs, our method significantly outperforms the HM reference codec of H.265 in its low-decay B configuration in terms of performance. Moreover, due to the lack of extensive code optimizations and GPU acceleration in the reference software, our method exhibits encoding speed far exceeding those of codecs on GPU, with decoding speed being comparable. In comparison to other neural video codecs, our method demonstrates performance close to that of DCVC-HEM, while on an RTX 2080Ti, our decoding speed is 7 times faster, and encoding speed is nearly 3 times faster.

\textbf{Quantitative Results.} Table \ref{tab1} and \ref{tab2} respectively show the BD-Rate (\%) comparison in terms of PSNR and MS-SSIM, using the low-delay B configuration of VTM as anchor. As shown in Table \ref{tab1}, our model exhibits only on average 18.9\% PSNR performance gap compared to the optimal VTM configuration, but has competitive performance compared to DCVC-HEM. From the perspective of MS-SSIM, Table \ref{tab2} reveals that our codec saves an average of 13.39\% bitrate across all datasets compared to VTM. Table \ref{tab3} supplements the BD-Rate (\%) comparison in terms of PSNR between our method and other methods under a different testing condition, using low-delay P configuration of VTM as anchor. Compared to the low-delay P configuration of VTM, our model exhibits better compression ratios on the UVG and MCL-JCV datasets. Additionally, our method demonstrates competitive or superior performance compared to other methods in Table \ref{tab3}. Table \ref{tab1} and Table \ref{tab_model_comparison} collectively demonstrate that our method achieves a new milestone in balancing rate-distortion performance with practical inference speed. With the future trend leaning towards ultra-high-definition video technology, the advantages of our method will become even more apparent.

As a supplement, to compare our approach with another study on efficient video coding \cite{mm2023}, due to the lack of open source code, we followed the testing condition outlined in the paper and tested the complexity and decoding speed of our model on 720P frames using a similar GPU. As shown in Table \ref{cmp_mm2023}, our method achieves a bitrate saving of 51.92\% without utilizing the OEU strategy. Although Tian\textit{ et al.} \cite{mm2023} employs model pruning while we do not, resulting in a higher overall complexity and slightly slower decoding speed for our method, the gap in decoding speed is reasonable given the performance improvements it brings forth.

\begin{table}[t]
\caption{Comparison of complexity and decoding time for 720p resolution inputs. Note that MM 2023 \cite{mm2023} employs model pruning, reducing the complexity of the P-frame model from the original 1.100T to 162G MACs.}
\centering
\resizebox{\linewidth}{!}{
\renewcommand{\arraystretch}{1.25}
\begin{tabular}{ccccc}
\toprule[1.0pt]
                                    & BD-rate(\%)  & MACs(Enc/Dec)          & Decoding time   & Device    \\ \hline
MM 2023 \cite{mm2023}                              & 0            & 162G/76G       & 40ms           & RTX 2080   \\  \hline
LRVC (ours)                                & -51.92       & 858G/130G       & 52ms           & RTX 2080Ti \\
\bottomrule[1.0pt]
\end{tabular}
}
\\
\vspace{-0.3cm}
\label{cmp_mm2023}
\end{table}

\subsection{Ablation Study}
\label{sec:abstudy}
In this section, we present ablation experiments on various strategies mentioned in this paper.

\textbf{Low-Resolution Representation Learning.} We investigated the impact of high-resolution spatial operations on model performance by adjusting the depth and width of the DCVC-DC \cite{dcvc_dc} model and retraining it. Specifically, to maintain low latency, we removed the post-processing module in the context decoder of DCVC-DC while retaining other high-resolution spatial operations. Additionally, we decreased the number of stacked modules and channels appropriately. However, even with these modifications, the decoding speed of this model still lagged behind our model. Upon training the adjusted model, we observed that it failed to converge, as depicted in Table \ref{abstudy}. This underscores the necessity of optimizing these high-resolution spatial operations through low-resolution representation learning.

\textbf{From Optical Flow to Deformable Convolution.} To investigate the impact of different alignment methods on model performance when reducing the resolution of contexts, we redesigned the motion decoder of the model and retrained the entire model. Specifically, we first decode the optical flow to the original resolution, then downsample it to match the resolution of the contexts and perform optical flow alignment. As shown in Table \ref{abstudy}, replacing the deformable alignment in the feature domain with optical flow alignment leads to significant performance degradation. This demonstrates that feature domain optical flow alignment performs poorly at low resolution, while our proposed progressive deformable alignment method exhibits superior performance.

\textbf{Multi-Frame Priors.} At the encoding end, we employ multi-frame priors for more accurate motion estimation. At the decoding end, by utilizing motion vectors and latent representations from multiple frames, we leverage multi-frame priors in low-resolution space through inexpensive computations for parameter prediction in the entropy model. To verify the effectiveness of multi-frame priors, we replace them with single-frame priors. As shown in Table \ref{abstudy}, multi-frame priors achieve larger bitrate saving compared to single-frame priors.

\textbf{Joint Training Strategy.} In Table \ref{abstudy}, we also investigate the impact of the joint training strategy on the results. Incorporating the I-frame model into the optimization objective results in an 18.95\% bitrate saving, demonstrating that joint training facilitates improved quality of propagated features between I-frames and P-frames and better rate-distortion trade-offs.

\begin{table}[t]
\caption{BD-rate comparison for different strategies.}
\centering
\resizebox{\linewidth}{!}{
\begin{tabular}{ccc}
\toprule[1.0pt]
Strategy                      & BD-rate(\%) & Decoding time \\ \hline
Ours                & 0              & 94ms       \\  \hline
Rescale DCVC-DC                & 5150.56              & 145ms       \\ \hline 
Ours w/o deformable convolution     & 89.31          & 79ms       \\  \hline
Ours w/o joint training      & 18.95                 & 94ms       \\  \hline
Ours w/o temporal prior      & 7.26                  & 92ms       \\ 
\bottomrule[1.0pt]
\end{tabular}}
\label{abstudy}
\end{table}

\begin{table}[t]
\centering
\caption{BD-rate and Encode time comparison for different OEU strategies on the V100 device. }
\label{oeu}
\resizebox{\linewidth}{!}{
\begin{tabular}{ccc} 
\toprule[1.0pt]
Strategy                                                                        & BD-rate(\%) & Encoding time                   \\ 
\hline
Ours                                                                            & 0           & 257ms                           \\ 
\hline
\begin{tabular}[c]{@{}c@{}}Ours+OEU\\(Downsample)\end{tabular}                  & -2.43       & 257ms + 154ms $\times$ $N$ Step   \\ 
\hline
\begin{tabular}[c]{@{}c@{}}Ours+OEU\\(Crop, update all frames)\end{tabular}     & -7.91       & 257ms + 1719ms $\times$ $N$ Step  \\ 
\hline
\begin{tabular}[c]{@{}c@{}}Ours+OEU\\(Crop, update first $T$ frames)\end{tabular} & -6.62       & 257ms + 437ms $\times$ $N$ Step   \\
\bottomrule[1.0pt]
\end{tabular}}
\end{table}

\textbf{OEU Strategy.} Table \ref{oeu} illustrates the impact of the online encoder update strategy on model performance. Due to memory constraints, we are unable to utilize original resolution frames for training high-resolution sequences. Simply downscaling input frames leads to marginal performance improvements while cropping them into patches results in better performance. For instance, for 1080p frames, we crop them into overlapping patches of size 768x448. The training sequence length is denoted as $T$, and the number of training steps as $N$. Based on experiments, we set $T$ to 8 and $N$ to 5. Furthermore, since the position within the GOP of each frame is fixed during testing, we found that updating only the first $T$ frames not only significantly reduces encoding time but also maintains good performance.

\textbf{Limitation.} While our method boasts strengths, it also has limitations. Throughout our work, despite employing a strategy of jointly training the I-frame model and P-frame model, we did not thoroughly investigate the correspondence between the Lagrange multipliers of these two models during training. Additionally, the rate-distortion curves in terms of MS-SSIM shown in Figure \ref{rd_curve12} indicate that our method performs poorer at higher bitrate ranges. Furthermore, our method falters on the HEVC B dataset, which contains sequences with complex motion, highlighting a weakness in handling intricate motion patterns. 

\section{Conclusion}
In this study, we introduce an efficiency-optimized framework for learned video compression that focuses on low-resolution representation learning. Our research has delved into strategies such as feature reuse, joint training, leveraging multi-frame priors, and online encoder updates. These strategies have not only enabled our framework to achieve an approximate compression ratio compared to the VTM but have also provided significant advantages in encoding and decoding speeds over other neural video codecs. Unlike traditional video codecs, neural video codecs eliminate the need for intricate code optimizations. While deploying current neural video codecs may pose challenges and considerations regarding power-constrained devices such as mobile phones and compatibility, it is evident that these efforts hold promise for making substantial contributions to the field of video coding.

\begin{acks}
To Robert, for the bagels and explaining CMYK and color spaces.
\end{acks}

\bibliographystyle{ACM-Reference-Format}
\bibliography{sample-base}










\end{document}